\documentclass[letterpaper,english,reprint, aps]{revtex4-1}
\usepackage[T1]{fontenc}
\usepackage[latin9]{inputenc}
\usepackage{amsmath}
\usepackage{amssymb}
\usepackage{graphicx}

\makeatletter

\pdfpageheight\paperheight
\pdfpagewidth\paperwidth

\makeatother

\usepackage{babel}
\begin{document}
\preprint{version 4.0 by amz}
\title{Experimentally verifying anti-Kibble-Zurek behavior in a quantum system
under noisy control field}
\author{Ming-Zhong Ai}
\author{Jin-Ming Cui}
\email{jmcui@ustc.edu.cn}

\author{Ran He}
\author{Zhong-Hua Qian}
\author{Xin-Xia Gao}
\author{Yun-Feng Huang}
\email{hyf@ustc.edu.cn}

\author{Chuan-Feng Li}
\email{cfli@ustc.edu.cn}

\author{Guang-Can Guo}
\affiliation{CAS Key Laboratory of Quantum Information, University of Science and
Technology of China, Hefei, 230026, People's Republic of China.}
\affiliation{CAS Center For Excellence in Quantum Information and Quantum Physics,
University of Science and Technology of China, Hefei, 230026, People's
Republic of China.}
\begin{abstract}
Kibble-Zurek mechanism (KZM) is a universal framework which could
in principle describe phase transition phenomenon in any system with
required symmetry properties. However, a conflicting observation termed
anti-KZ behavior has been reported in the study of ferroelectric phase
transition, in which slower driving results in more topological defects
{[}S. M. Griffin, et al. Phys. Rev. X. 2, 041022 (2012){]}. Although
this research is significant, its experimental simulations have been
scarce until now. In this work, we experimentally demonstrate anti-KZ
behavior under noisy control field in three kinds of quantum phase
transition protocols using a single trapped ${\rm ^{171}Yb^{+}}$
ion. The density of defects is studied as a function of the quench
time and the noise intensity. We experimentally verify that the optimal
quench time to minimize excitation scales as a universal power law
of the noise intensity. Our research sets a stage for quantum simulation
of such anti-KZ behavior in two-level systems and reveals the limitations
of the adiabatic protocols such as quantum annealing.
\end{abstract}
\maketitle
Kibble-Zurek mechanism (KZM), which is originally proposed to describe
early-universe phase transition by Kibble and Zurek \citep{kibble1976topology,zurek1985cosmological},
provides an elegant theoretical framework for exploring the critical
dynamics of phase transition \citep{campo2014universality}. Its central
prediction is that the density of topological defects $n_{0}$, formed
when a system is driven through a critical point in a time scale $\tau$,
follows a universal power law as a function of quench time: $n_{0}\propto\tau^{-\beta}$.
The power-law exponent $\beta=d\nu/(1+z\nu)>0$ is determined by the
dimensionality of the system $d$, equilibrium correlation-length
$\nu$ and dynamic critical exponents $z$ respectively \citep{dutta2016anti}.
Notably, in the quantum domain, the KZM provides useful heuristic
for the preparation of ground-state phases of matter in quantum simulation
as well as for adiabatic quantum computation \citep{suzuki2010quench}.
Although the KZM has many important implications, its experimental
verification still calls for further studies. For classical continuous
phase transitions, many systems have verified this mechanism, such
as cold atomic gases \citep{navon2015critical}, ion crystals \citep{ulm2013observation,pyka2013topological},
and superconductors \citep{monaco2001dynamics}. Meanwhile for quantum
phase transitions, which are accomplished by varying a parameter in
the Hamiltonian in order to tune between different quantum phases,
its experimental verification are still scarce due to the difficulty
of exactly controlling driven parameters \citep{chen2011quantum,braun2015emergence,anquez2016quantum,gardas2018defects,keesling2018probing}.
And it had been performed only in few platforms through quantum simulators
\citep{xu2014quantum,cui2016j,gong2016m,cui2020experimentally}.

\begin{figure}
\includegraphics{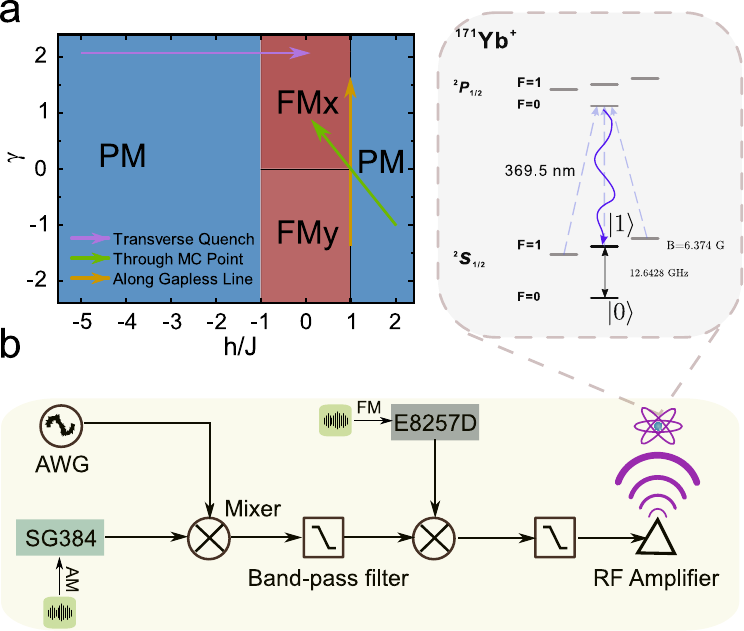}

\caption{\label{fig:Fig-1} (color online). Phase diagram of three quantum
phase transition protocols and the schematic diagram of the experimental
device. (a) The phase diagram is divided into para magnetic phase
and ferromagnetic phase which are denoted by PM and FM respectively.
These two phase is separated by the parameter $h/J=\pm1$ in our $\gamma-hJ$
frame. The middle ferromagnetic phase is also divided into two parts
by the line $\gamma=0$, which ordering along $x$ and $y$ directions.
The three lines with arrow represent three quench protocols explained
in legend. (b) The microwave used in our experiments is generated
by a mixing wave scheme. The illustration in (a) is the energy level
diagram of $^{171}{\rm Yb}^{+}$ ion. }
\end{figure}

\begin{figure*}
\includegraphics[width=15cm]{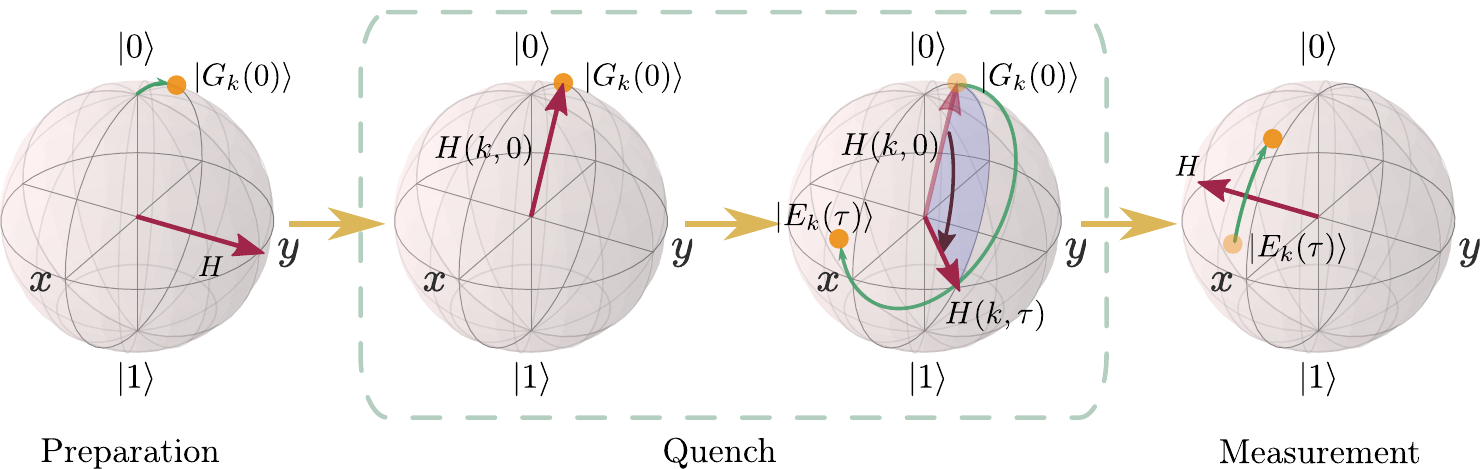}\caption{\label{fig:Fig_opera} (color online). Scheme to measure the excitation
probability. The quantum critical dynamics of the one-dimensional
transverse-field XY chain model is detected by measuring corresponding
Landau\textendash Zener crossings dynamics in each mode. For each
mode, a typical process to measure the excitation probability includes
preparation, quench and measurement.}
\end{figure*}

While the KZM is believed to be broadly applicable, a conflicting
observation has been reported in the study of ferroelectric phase
transition: slower quenches generate more topological defects when
approaching the adiabatic limit. Opposite to that predicted by standard
KZM, this counter intuitive phenomenon is termed as anti-Kibble-Zurek
(anti-KZ) dynamics \citep{griffin2012scaling}. Considerable attention
has been devoted to the anti-KZ mechanism in the last few years. The
universal properties of quantum quenches of a quantum system coupling
to thermal dissipation simulated using transverse field Ising model
is theoretically studied in \citep{patane2008adiabatic,nalbach2015quantum},
which exhibits anti-KZ behavior. Meanwhile, Adolfo et al. show a thermally
isolated system driven across a quantum phase transition under a noisy
control field also exhibits anti-KZ behavior, whereby slower driving
results in higher density of defects \citep{dutta2016anti}. In order
to explore whether the anti-KZ behavior will appear in other quantum
spin models with different scaling exponents under noisy control fields,
dynamics of a transverse-field XY chain driven across quantum critical
points under noisy control fields is theoretically studied in \citep{gao2017anti}.
However, previous studies are just present theoretically or investigated
under unwanted or uncontrolled noisy fields \citep{dutta2016anti,puebla2020universal,griffin2012scaling,liou2018quench,cui2020experimentally}.
Unfortunately, those experiments are too limited to verify the the
most important scaling behavior predict by the theory.

In this paper, we experimentally investigated anti-KZ mechanism in
quantum phase transition by applying fully controlled noisy driving
fields on the two level system (TLS) with Landau-Zener (LZ) crossings
in a trapped ${\rm ^{171}Yb^{+}}$ ion, and clearly verified three
different scaling exponents of the transverse-field XY chain. Different
scaling exponents are realized through quenching across the boundary
line between para magnetic and ferromagnetic phase, quenching across
the isolated multicritical (MC) point and quenching along the gapless
line, respectively \citep{mukherjee2007v,divakaran2009defect,divakaran2008u}.
Thanks to the precise control of Gaussian noise, we quantitatively
investigate the density of topological defects as a function of the
quench time and the intensity of Gaussian noise. The results agree
well with the theoretical expectation in \citep{gao2017anti}, in
which the optimal quench time to minimize defects scales as a universal
power law of the noise intensity in all three protocols.

\begin{figure*}
\includegraphics[width=17cm]{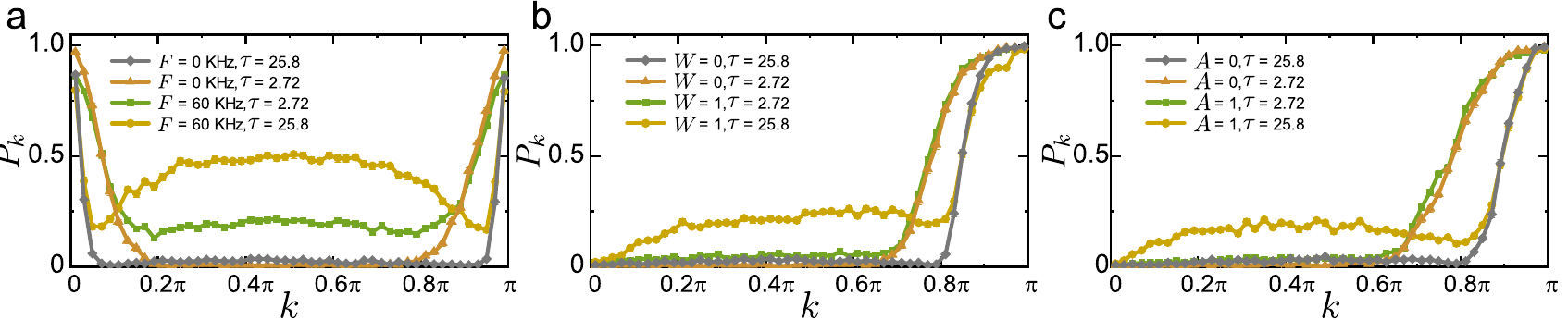}\caption{\label{fig:Fig2} (color online). The probabilities of excited state
$p_{k}$ as a function of mode $k$ for all three quench protocols.
(a) Transverse quench protocol. The ginger and green symbols represent
probabilities under noisy control field with frequency deviation 60
KHz. The white Gaussian noise causes excitation near $k=\pi/2$ while
the noise-free cases denoted by brown and gray symbols are very flat.
(b) and (c) are quenching along gapless line and through MC point
protocols respectively. Every protocol is demonstrated with and without
noise in control field. For each point, the experiment is repeated
1000 times and the error bars indicate a standard deviation.}
\end{figure*}

According to Jordan-Wigner (J-W) transformation \citep{lieb1961two,bunder1999je,caneva2007adiabatic},
a spin-1/2 quantum transverse field XY chain Hamiltonian can be transformed
to the fermion form:

\begin{equation}
\begin{aligned}H_{1}= & -J\sum_{l=1}^{N}[(c_{l}^{\dagger}c_{l+1}+c_{l+1}^{\dagger}c_{l})+\gamma(c_{l}^{\dagger}c_{l+1}^{\dagger}+c_{l+1}c_{l})]\\
 & -h\sum_{l=1}^{N}(2c_{l}^{\dagger}c_{l}-1).
\end{aligned}
\end{equation}
in which $c_{l}$ is obtained by J-W transformation. The variable
$N$ counts the number of spins, $h$ measures the strength of the
transverse field. We set $J=J_{x}+J_{y}$, $\gamma=(J_{x}-J_{y})/J$
where $J_{x}$ and $J_{y}$ represent the anisotropy interactions
along $x$ and $y$ spin directions respectively. This Hamiltonian
can be decoupled into a sum of independent terms $H_{1}=\sum_{k\in[0,\pi]}H_{m}(k)$,
where the Hamiltonian density $H_{m}(k)$ in pseodumomentum space
can be written as:

\begin{equation}
H_{m}(k)=-2[\sigma_{z}(J{\rm cos}k+h)+\sigma_{x}(J\gamma{\rm sin}k)],\label{eq:pse_H}
\end{equation}
in which $\sigma_{z}$ and $\sigma_{x}$ are Pauli matrices. The evolution
of the generic state $\psi_{k}(t)$ is governed by Schr\"odinger
equation $i\frac{d}{dt}\left|\psi_{k}(t)\right\rangle =H_{m}(k,t)\left|\psi_{k}(t)\right\rangle $.
This reduces the quantum many-body transverse field XY chain Hamiltonian
to an array of decoupled single spin-1/2 Hamiltonian, which could
be simulated utilizing a TLS with well-designed Landau-Zener crossings
experimentally, such as a trapped ion qubit.

For the convenience of experimental demonstration, variation of one
parameter in $H_{m}(k)$ is considered. The phase diagram of the transverse-field
XY chain, which is spanned by parameters $h/J$ and $\gamma$, is
divided into four parts: the quantum paramagnetic phase PM and two
ferromagnetic long-ranged phases ordering along x and y directions
denoted by FMx and FMy respectively as shown in Fig. \ref{fig:Fig-1}(a).
The definition of the density of defects in this transverse field
XY chain after quench is similar to the case for the Ising model \citep{dziarmaga2010dynamics,dziarmaga2005dynamics,zurek2005dynamics},
which could be denoted by:

\begin{equation}
n_{W}=\frac{1}{N_{k}}\sum_{k\in[0,\pi]}p_{k},\label{eq:n_w}
\end{equation}
where $N_{k}$ is the number of $k$-modes used in the summation of
Hamiltonian $H_{m}(k)$, and $p_{k}$ is the probability measured
in the excited state $\left|E_{k}(\tau)\right\rangle $ after evolution
driven under the $k$th mode Hamiltonian from $\left|G_{k}(0)\right\rangle $.
Notably, $\{\left|G_{k}(t)\right\rangle ,\left|E_{k}(t)\right\rangle \}$
is the basis of adiabatic instantaneous eigenstate of $H_{m}(k)$.

In order to observe anti-KZ phenomenon, driving with noisy control
fields in the simulation is considered. White Gaussian noise is a
good approximation to ubiquitous colored noise with exponentially
decaying correlations, therefore the noise term $\eta(t)$ is set
as white Gaussian noise with zero mean and second moment $\overline{\eta(t)\eta(t')}=W^{2}\delta(t-t')$.
Here $W^{2}$ represents the intensity of the noise fluctuation. We
add this noise term to quench parameter in the form of $f(t)=f^{(0)}(t)+\eta(t)$,
where $f^{(0)}(t)\propto t/\tau$ is the perfect control parameter
linearly varying in time with quench time $\tau$. The noise control
fields in our experiments are produced by mixing method. The intensity
of this noise can be accurately controlled by the modulation parameters.
This method can also be applied to other quantum simulation experiments
and open up a new way for quantum simulation experiment with noise.

Our experiments are performed using a trapped ${\rm ^{171}Yb^{+}}$
ion in needle trap with the setup simplified shown in Fig. \ref{fig:Fig-1}(b).
Two hyperfine levels of $^{171}{\rm Yb}^{+}$ ion in the $S_{1/2}$
ground state, which means $\left|^{2}S_{1/2},F=0,m_{F}=0\right\rangle $
and $\left|^{2}S_{1/2},F=1,m_{F}=0\right\rangle $, are encoded to
$\left|0\right\rangle $ and $\left|1\right\rangle $ respectively.
The microwave used to drive this ion qubit is generated through mixing
twice. A microwave signal around 200 MHz generated from a two channel
Arbitrary Waveform Generator (AWG) is mixed with a 3.0 GHz microwave
generated from a RF signal generator (SG384, Stanford Research Systems).
This mixed signal is mixed again with a 9.44 GHz microwave generated
from a Analog Signal Generator (E8257D, Agilent) to obtain an arbitrary
microwave near 12.64 GHz, and then this signal is amplified to about
2 W and irradiated to the trapped ion by a horn antenna. In all of
our experiments, the Rabi time is set to 100 $\mu{\rm s}$ and all
expressions of $\tau$ in the following text represent multiples of
the Rabi time.

We first consider the transverse quench, in which case only the parameter
$h(t)$ is time-dependent, as shown in Fig. \ref{fig:Fig-1}(a). To
simulate the quench dynamics under noise fluctuation, white Gaussian
noise $\eta(t)$ is added to the time-dependent quench parameter $h(t)$
as described above. The Hamiltonian of Eq. \ref{eq:pse_H} can be
rewritten as:

\begin{equation}
\begin{aligned}H_{m}^{(1)}(k,t)= & -2[(J_{x}+J_{y}){\rm cos}k+h(t)]\sigma_{z}\\
 & -2[(J_{x}-J_{y}){\rm sin}k]\sigma_{x}-2\eta(t)\sigma_{z}.
\end{aligned}
\label{eq:protocol 1}
\end{equation}
This Hamiltonian can be transformed into standard LZ model $H_{LZ}(k,t)=-\frac{1}{2}(\sigma_{x}+\nu_{LZ}t_{LZ}\sigma_{z})$
using the substitutions $\nu_{LZ}=\nu_{h}/(2J\gamma{\rm sin}k)^{2},t_{LZ}=4J\gamma{\rm sin}k(t+J{\rm cos}k/\nu_{h})$,
in which $h(t)=\nu_{h}t$ and $\nu_{h}$ is the quench velocity. The
standard LZ model could be simulated through a trapped ion qubit as
described in Ref. \citep{cui2016j}. We first drive the qubit to the
ground-state $\left|G_{k}(0)\right\rangle $ of Hamiltonian $H_{m}^{(1)}(k,0)$
from initial state $\left|0\right\rangle $. Then the simulator will
evolve under the control of this Hamiltonian. The quench parameter
$h(t)$ varies linearly from -5 to 0 with entire quench time $\tau$
while the other two independent parameters are fixed as $J_{x}=1$
and $J_{y}=-1/3$ in the evolution. Finally the state is driven again
to the basis $\{\left|0\right\rangle ,\left|1\right\rangle \}$, which
is the reverse process of the first process, to measure the population
probability $p_{k}$ of the excited state $\left|E_{k}(\tau)\right\rangle $
by fluorescence detection scheme. The white Gaussian noise in this
quench protocol is generated through frequency modulation (FM) the
microwave generated by SG384 utilizing built-in noise source (the
detailed form of this noise is described in Supplemental Material).
Different noise intensity are realized through varying frequency deviation
$F$ in FM (explained in the Supplemental Material). We decompose
the Hamiltonian $H(t)$ into 50 independent terms $H_{m}(k,t)$ in
all three protocols, and the parameter $k$ is sampled 50 times equidistantly
from 0 to $\pi$. The final population probability $p_{k}$ as a function
of $k$ under different noise intensity and quench time are shown
in Fig. \ref{fig:Fig2}(a). As a result, the white Gaussian noise
causes bulge around $k=\pi/2$, which is the reason of addition of
the density of defects in this quench process. And the stronger the
noise is, the more defects would be generated.

\begin{figure}
\includegraphics{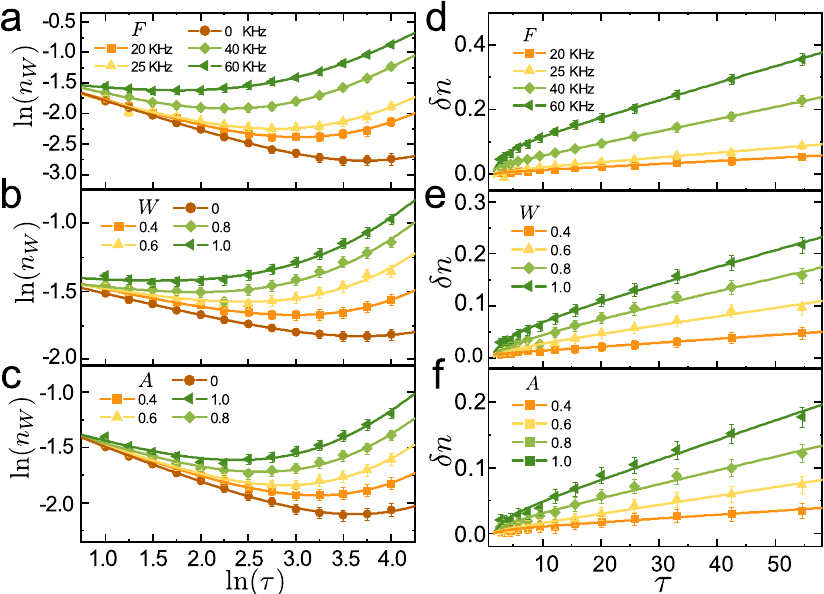}\caption{\label{fig:Fig3} (color online). The anti-KZ behavior of defects
production in three quench protocols. The defects density ${\rm ln}(n_{W})$
as a function of quench time ${\rm ln}(\tau)$ are shown in (a), (b)
and (c), in which the defects decrease first and then increase in
the limitation of long quench time. At the same quench time, stronger
noise cause more defects. Due to the existence of system noise, it
will also demonstrate anti-KZ behavior even if the Gaussian noise
is zero. The corresponding Gaussian noise-induced defect densities
$\delta n=n_{W}-n_{0}$ in these three cases are shown in (d), (e)
and (f), which are proportional to quench time $\tau$. These picture's
layout are arranged in the order of transverse quench, anisotropic
quench across the multicritical point and quench along the gapless
line. For each point, the experiment is repeated 1000 times and the
error bars indicate a standard deviation.}
\end{figure}

We proceed to consider the second quench protocol, the anisotropic
quench across the multicritical point, as shown in Fig. \ref{fig:Fig-1}(b).
The Hamiltonian for each $k$-mode in this case can be rewritten as:

\begin{equation}
\begin{aligned}H_{m}^{(2)}(k,t)= & -2\{[J_{x}(t)+J_{y}]{\rm cos}(k)+h\}\sigma_{z}\\
 & -2[(J_{x}(t)-J_{y}){\rm sin}(k)]\sigma_{x}\\
 & -2{\eta(t)[({\rm sin}k)\sigma_{x}+({\rm cos}k)\sigma_{z}]}
\end{aligned}
\end{equation}
with time-dependent quench parameter $J_{x}(t)$ ramping from 0 to
3. The Hamiltonian $H_{m}^{(2)}(k,t)$ can be transformed into standard
LZ model using the substitutions $\nu_{LZ}=\nu_{x}/[2(J_{y}{\rm sin}2k+h{\rm sin}k)]^{2},t_{LZ}=4(J_{y}{\rm sin}2k+h{\rm sin}k)\times[t+(J_{y}{\rm cos}2k+h{\rm cos}k)/\nu_{x}]$.
Similar to the first protocol, we fix $h=2$ and $J_{y}=1$ in all
experiments of this protocol. Under this condition, the system is
initially in the PM phase and then is driven through the multicritical
point into the FMx phase. The noise used in this quench protocol is
induced through frequency modulation (FM) SG384 and amplitude modulation
(AM) E8257D synchronously utilizing built-in Gaussian noise respectively.
Different noise intensity is realized by changing frequency deviation
$F$ and modulation depth $A$ proportionally (explained in the Supplemental
Material). For convenience, we represent the noise intensity $W^{2}$
through percentage, which in AM is the modulation depth $W^{2}=A^{2}$
and in FM is the ratio $W^{2}=(F/60{\rm kHz)^{2}}$. The probabilities
$p_{k}$ measured in excited state as a function of $k$ with different
noise intensity and quench time are shown in Fig. \ref{fig:Fig2}(b). 

\begin{figure}
\includegraphics{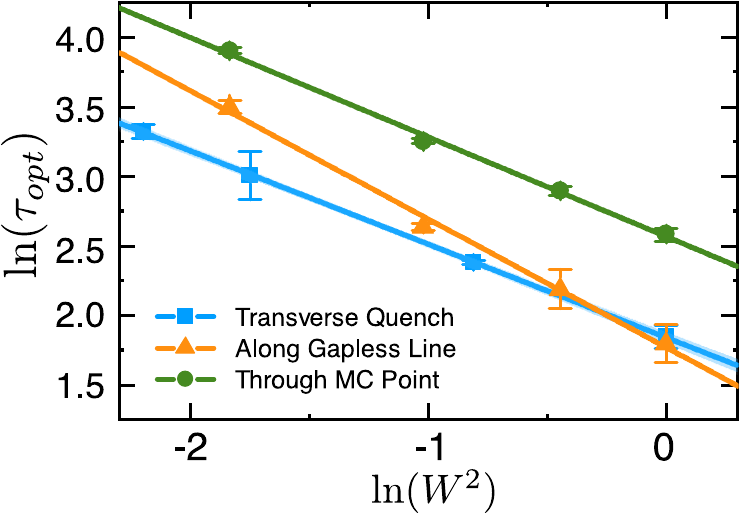}\caption{\label{fig:Fig4} (color online). The linear relationship of logarithm
of optimal quench time ${\rm ln}(\tau_{opt})$ as a function of logarithm
of noise intensity ${\rm ln}(W^{2})$ gives ${\rm ln}(\tau_{opt})\propto\alpha_{fit}{\rm ln}(W^{2})$.
The fitting parameters of these three quench protocols are: $\alpha_{{\rm fit}}^{1}=-0.67\pm0.01$
for the transverse quench, which is close to analytical result $\alpha_{{\rm theory}}^{1}=-0.67$;
$\alpha_{{\rm fit}}^{2}=-0.92\pm0.05$ for the quench through the
multicritical point with $\alpha_{{\rm theory}}^{2}=-0.86$; $\alpha_{{\rm fit}}^{3}=-0.71\pm0.03$
for the quench along gapless line with $\alpha_{{\rm theory}}=-0.75$.
The error bars indicate a standard deviation.}
\end{figure}

For the last quench protocol along the gapless line, the Hamiltonian
for each $k$-mode is:

\begin{equation}
\begin{aligned}H_{m}^{(3)}(k,t)= & -2[(J{\rm cos}k+h)\sigma_{z}+J\gamma(t){\rm sin}k\sigma_{x}]\\
 & -2\eta(t)J{\rm sin}k\sigma_{x},
\end{aligned}
\end{equation}
in which the time-dependent parameter $\gamma(t)$ ramps from -2 to
2 while the other parameters are fixed as $h=1$ and $J=J_{x}+J_{y}=1$.
This Hamiltonian could be transformed into standard LZ model using
the substitutions $\nu_{LZ}=\nu_{\gamma}{\rm sin}k/[2({\rm cos}k+1)]^{2},t_{LZ}=-4({\rm cos}k+1)t$.
The noise is induced through amplitude modulation (AM) the microwave
source E8257D utilizing built-in Gaussian noise. Figure \ref{fig:Fig2}(c)
shows the probabilities $p_{k}$ measured in excited state as a function
of $k$ with different noise intensity and quench time. 

The defects density exhibits anti-KZ behavior in all of these three
quench protocols when the noise presents. This makes it possible to
find an optimal quench time $\tau_{opt}$ to minimize the defects
density. The defects density under the control of noise $n_{W}\approx r_{W}\tau+c\tau^{-\beta}$,
where the prefactor $c$ is predicted by KZM and $r_{W}$ characterizes
the intensity of the noise. Then we can define 

\begin{equation}
\delta n=n_{W}-n_{0}\approx r_{W}\tau+c\tau^{-\beta}-(r_{0}\tau+c\tau^{-\beta})=\delta r\tau
\end{equation}
to represent the defect density induced by noise in control field.
The results are shown in Figure \ref{fig:Fig3}. Now that the parameter
$r_{W}$ represents the productivity of defects under noisy control
field, we can remove the system noise by subtract $r_{0}$ from $r_{W}$
to indicate the efficiency of defects induced by noise in control
field, in which $r_{0}$ is the fitting parameter in noise-free driving
field in these protocols respectively. We use the expression $n_{W}\approx\delta r\tau+c\tau^{-\beta}$
to find the optimal quench time $\tau_{opt}$ to minimize $n_{W}$.
As illustrated in Fig. \ref{fig:Fig4}, the optimal quench time to
minimize defects scales as a power law of the noise intensity $W^{2}$
in all of these three protocols. Linear fitting ${\rm ln}(\tau_{opt})$
as a function of ${\rm ln}(W^{2})$ gives ${\rm ln}(\tau_{opt})\propto\alpha_{fit}{\rm ln}(W^{2})$
where the fitting parameters for the three cases agree well with analytical
result: $\alpha_{{\rm fit}}^{1}=-0.67\pm0.01$ for the transverse
quench with $\alpha_{{\rm theory}}^{1}=-2/3=0.67$; $\alpha_{{\rm fit}}^{2}=-0.92\pm0.05$
for the quench through the multicritical point with $\alpha_{{\rm theory}}^{2}=-6/7=-0.86$;
$\alpha_{{\rm fit}}^{3}=-0.71\pm0.03$ for the quench along gapless
line with $\alpha_{{\rm theory}}^{3}=-3/4=-0.75$ \citep{mukherjee2007quenching,divakaran2009defect,divakaran2008quenching}.

In summary, we for the first time experimentally studied the anti-KZ
behavior in three quantum phase transition protocols under white Gaussian
noisy control field using a single trapped ion. The experimental results
can be used as a powerful evidence for anti-KZ phenomenon. We also
show the optimal quench time to minimize defects density $\tau_{opt}$
scales as a universal power law of the noise intensity $W^{2}$ for
all of these three cases, which may inspire us about the limitations
of adiabatic protocols such as quantum annealing.
\begin{acknowledgments}
This work was supported by the National Key Research and Development
Program of China (Nos. 2017YFA0304100, 2016YFA0302700), the National
Natural Science Foundation of China (Nos. 11874343, 61327901, 11774335,
11474270, 11734015, 11874343), Key Research Program of Frontier Sciences,
CAS (No. QYZDY-SSW-SLH003), the Fundamental Research Funds for the
Central Universities (Nos. WK2470000026, WK2470000018), An-hui Initiative
in Quantum Information Technologies (AHY020100, AHY070000), the National
Program for Support of Topnotch Young Professionals (Grant No. BB2470000005). 
\end{acknowledgments}

\bibliographystyle{apsrev4-1}
\bibliography{antiKZM}

\begin{thebibliography}{35}%
\makeatletter
\providecommand \@ifxundefined [1]{%
 \@ifx{#1\undefined}
}%
\providecommand \@ifnum [1]{%
 \ifnum #1\expandafter \@firstoftwo
 \else \expandafter \@secondoftwo
 \fi
}%
\providecommand \@ifx [1]{%
 \ifx #1\expandafter \@firstoftwo
 \else \expandafter \@secondoftwo
 \fi
}%
\providecommand \natexlab [1]{#1}%
\providecommand \enquote  [1]{``#1''}%
\providecommand \bibnamefont  [1]{#1}%
\providecommand \bibfnamefont [1]{#1}%
\providecommand \citenamefont [1]{#1}%
\providecommand \href@noop [0]{\@secondoftwo}%
\providecommand \href [0]{\begingroup \@sanitize@url \@href}%
\providecommand \@href[1]{\@@startlink{#1}\@@href}%
\providecommand \@@href[1]{\endgroup#1\@@endlink}%
\providecommand \@sanitize@url [0]{\catcode `\\12\catcode `\$12\catcode
  `\&12\catcode `\#12\catcode `\^12\catcode `\_12\catcode `\%12\relax}%
\providecommand \@@startlink[1]{}%
\providecommand \@@endlink[0]{}%
\providecommand \url  [0]{\begingroup\@sanitize@url \@url }%
\providecommand \@url [1]{\endgroup\@href {#1}{\urlprefix }}%
\providecommand \urlprefix  [0]{URL }%
\providecommand \Eprint [0]{\href }%
\providecommand \doibase [0]{http://dx.doi.org/}%
\providecommand \selectlanguage [0]{\@gobble}%
\providecommand \bibinfo  [0]{\@secondoftwo}%
\providecommand \bibfield  [0]{\@secondoftwo}%
\providecommand \translation [1]{[#1]}%
\providecommand \BibitemOpen [0]{}%
\providecommand \bibitemStop [0]{}%
\providecommand \bibitemNoStop [0]{.\EOS\space}%
\providecommand \EOS [0]{\spacefactor3000\relax}%
\providecommand \BibitemShut  [1]{\csname bibitem#1\endcsname}%
\let\auto@bib@innerbib\@empty
\bibitem [{\citenamefont {Kibble}(1976)}]{kibble1976topology}%
  \BibitemOpen
  \bibfield  {author} {\bibinfo {author} {\bibfnamefont {T.~W.}\ \bibnamefont
  {Kibble}},\ }\href@noop {} {\bibfield  {journal} {\bibinfo  {journal}
  {Journal of Physics A: Mathematical and General}\ }\textbf {\bibinfo {volume}
  {9}},\ \bibinfo {pages} {1387} (\bibinfo {year} {1976})}\BibitemShut
  {NoStop}%
\bibitem [{\citenamefont {Zurek}(1985)}]{zurek1985cosmological}%
  \BibitemOpen
  \bibfield  {author} {\bibinfo {author} {\bibfnamefont {W.~H.}\ \bibnamefont
  {Zurek}},\ }\href@noop {} {\bibfield  {journal} {\bibinfo  {journal}
  {Nature}\ }\textbf {\bibinfo {volume} {317}},\ \bibinfo {pages} {505}
  (\bibinfo {year} {1985})}\BibitemShut {NoStop}%
\bibitem [{\citenamefont {CAMPO}\ and\ \citenamefont
  {Zurek}(2014)}]{campo2014universality}%
  \BibitemOpen
  \bibfield  {author} {\bibinfo {author} {\bibfnamefont {A.~D.}\ \bibnamefont
  {CAMPO}}\ and\ \bibinfo {author} {\bibfnamefont {W.~H.}\ \bibnamefont
  {Zurek}},\ }in\ \href@noop {} {\emph {\bibinfo {booktitle} {Symmetry and
  Fundamental Physics: Tom Kibble at 80}}}\ (\bibinfo  {publisher} {World
  Scientific},\ \bibinfo {year} {2014})\ pp.\ \bibinfo {pages}
  {31--87}\BibitemShut {NoStop}%
\bibitem [{\citenamefont {Dutta}\ \emph {et~al.}(2016)\citenamefont {Dutta},
  \citenamefont {Rahmani},\ and\ \citenamefont {del Campo}}]{dutta2016anti}%
  \BibitemOpen
  \bibfield  {author} {\bibinfo {author} {\bibfnamefont {A.}~\bibnamefont
  {Dutta}}, \bibinfo {author} {\bibfnamefont {A.}~\bibnamefont {Rahmani}}, \
  and\ \bibinfo {author} {\bibfnamefont {A.}~\bibnamefont {del Campo}},\
  }\href@noop {} {\bibfield  {journal} {\bibinfo  {journal} {Physical review
  letters}\ }\textbf {\bibinfo {volume} {117}},\ \bibinfo {pages} {080402}
  (\bibinfo {year} {2016})}\BibitemShut {NoStop}%
\bibitem [{\citenamefont {Suzuki}(2010)}]{suzuki2010quench}%
  \BibitemOpen
  \bibfield  {author} {\bibinfo {author} {\bibfnamefont {S.}~\bibnamefont
  {Suzuki}},\ }in\ \href@noop {} {\emph {\bibinfo {booktitle} {Quantum
  Quenching, Annealing and Computation}}}\ (\bibinfo  {publisher} {Springer},\
  \bibinfo {year} {2010})\ pp.\ \bibinfo {pages} {115--143}\BibitemShut
  {NoStop}%
\bibitem [{\citenamefont {Navon}\ \emph {et~al.}(2015)\citenamefont {Navon},
  \citenamefont {Gaunt}, \citenamefont {Smith},\ and\ \citenamefont
  {Hadzibabic}}]{navon2015critical}%
  \BibitemOpen
  \bibfield  {author} {\bibinfo {author} {\bibfnamefont {N.}~\bibnamefont
  {Navon}}, \bibinfo {author} {\bibfnamefont {A.~L.}\ \bibnamefont {Gaunt}},
  \bibinfo {author} {\bibfnamefont {R.~P.}\ \bibnamefont {Smith}}, \ and\
  \bibinfo {author} {\bibfnamefont {Z.}~\bibnamefont {Hadzibabic}},\
  }\href@noop {} {\bibfield  {journal} {\bibinfo  {journal} {Science}\ }\textbf
  {\bibinfo {volume} {347}},\ \bibinfo {pages} {167} (\bibinfo {year}
  {2015})}\BibitemShut {NoStop}%
\bibitem [{\citenamefont {Ulm}\ \emph {et~al.}(2013)\citenamefont {Ulm},
  \citenamefont {Ro{\ss}nagel}, \citenamefont {Jacob}, \citenamefont
  {Deg{\"u}nther}, \citenamefont {Dawkins}, \citenamefont {Poschinger},
  \citenamefont {Nigmatullin}, \citenamefont {Retzker}, \citenamefont {Plenio},
  \citenamefont {Schmidt-Kaler} \emph {et~al.}}]{ulm2013observation}%
  \BibitemOpen
  \bibfield  {author} {\bibinfo {author} {\bibfnamefont {S.}~\bibnamefont
  {Ulm}}, \bibinfo {author} {\bibfnamefont {J.}~\bibnamefont {Ro{\ss}nagel}},
  \bibinfo {author} {\bibfnamefont {G.}~\bibnamefont {Jacob}}, \bibinfo
  {author} {\bibfnamefont {C.}~\bibnamefont {Deg{\"u}nther}}, \bibinfo {author}
  {\bibfnamefont {S.}~\bibnamefont {Dawkins}}, \bibinfo {author} {\bibfnamefont
  {U.}~\bibnamefont {Poschinger}}, \bibinfo {author} {\bibfnamefont
  {R.}~\bibnamefont {Nigmatullin}}, \bibinfo {author} {\bibfnamefont
  {A.}~\bibnamefont {Retzker}}, \bibinfo {author} {\bibfnamefont
  {M.}~\bibnamefont {Plenio}}, \bibinfo {author} {\bibfnamefont
  {F.}~\bibnamefont {Schmidt-Kaler}},  \emph {et~al.},\ }\href@noop {}
  {\bibfield  {journal} {\bibinfo  {journal} {Nature communications}\ }\textbf
  {\bibinfo {volume} {4}},\ \bibinfo {pages} {2290} (\bibinfo {year}
  {2013})}\BibitemShut {NoStop}%
\bibitem [{\citenamefont {Pyka}\ \emph {et~al.}(2013)\citenamefont {Pyka},
  \citenamefont {Keller}, \citenamefont {Partner}, \citenamefont {Nigmatullin},
  \citenamefont {Burgermeister}, \citenamefont {Meier}, \citenamefont
  {Kuhlmann}, \citenamefont {Retzker}, \citenamefont {Plenio}, \citenamefont
  {Zurek} \emph {et~al.}}]{pyka2013topological}%
  \BibitemOpen
  \bibfield  {author} {\bibinfo {author} {\bibfnamefont {K.}~\bibnamefont
  {Pyka}}, \bibinfo {author} {\bibfnamefont {J.}~\bibnamefont {Keller}},
  \bibinfo {author} {\bibfnamefont {H.}~\bibnamefont {Partner}}, \bibinfo
  {author} {\bibfnamefont {R.}~\bibnamefont {Nigmatullin}}, \bibinfo {author}
  {\bibfnamefont {T.}~\bibnamefont {Burgermeister}}, \bibinfo {author}
  {\bibfnamefont {D.}~\bibnamefont {Meier}}, \bibinfo {author} {\bibfnamefont
  {K.}~\bibnamefont {Kuhlmann}}, \bibinfo {author} {\bibfnamefont
  {A.}~\bibnamefont {Retzker}}, \bibinfo {author} {\bibfnamefont {M.~B.}\
  \bibnamefont {Plenio}}, \bibinfo {author} {\bibfnamefont {W.}~\bibnamefont
  {Zurek}},  \emph {et~al.},\ }\href@noop {} {\bibfield  {journal} {\bibinfo
  {journal} {Nature communications}\ }\textbf {\bibinfo {volume} {4}},\
  \bibinfo {pages} {2291} (\bibinfo {year} {2013})}\BibitemShut {NoStop}%
\bibitem [{\citenamefont {Monaco}\ \emph {et~al.}(2001)\citenamefont {Monaco},
  \citenamefont {Rivers},\ and\ \citenamefont {Mygind}}]{monaco2001dynamics}%
  \BibitemOpen
  \bibfield  {author} {\bibinfo {author} {\bibfnamefont {R.}~\bibnamefont
  {Monaco}}, \bibinfo {author} {\bibfnamefont {R.}~\bibnamefont {Rivers}}, \
  and\ \bibinfo {author} {\bibfnamefont {J.}~\bibnamefont {Mygind}},\
  }\href@noop {} {\emph {\bibinfo {title} {The Dynamics of Spontaneous Fluxon
  formation in Annular Josephson Tunnel Junctions}}},\ \bibinfo {type} {Tech.
  Rep.}\ (\bibinfo {year} {2001})\BibitemShut {NoStop}%
\bibitem [{\citenamefont {Chen}\ \emph {et~al.}(2011)\citenamefont {Chen},
  \citenamefont {White}, \citenamefont {Borries},\ and\ \citenamefont
  {DeMarco}}]{chen2011quantum}%
  \BibitemOpen
  \bibfield  {author} {\bibinfo {author} {\bibfnamefont {D.}~\bibnamefont
  {Chen}}, \bibinfo {author} {\bibfnamefont {M.}~\bibnamefont {White}},
  \bibinfo {author} {\bibfnamefont {C.}~\bibnamefont {Borries}}, \ and\
  \bibinfo {author} {\bibfnamefont {B.}~\bibnamefont {DeMarco}},\ }\href@noop
  {} {\bibfield  {journal} {\bibinfo  {journal} {Physical Review Letters}\
  }\textbf {\bibinfo {volume} {106}},\ \bibinfo {pages} {235304} (\bibinfo
  {year} {2011})}\BibitemShut {NoStop}%
\bibitem [{\citenamefont {Braun}\ \emph {et~al.}(2015)\citenamefont {Braun},
  \citenamefont {Friesdorf}, \citenamefont {Hodgman}, \citenamefont
  {Schreiber}, \citenamefont {Ronzheimer}, \citenamefont {Riera}, \citenamefont
  {Del~Rey}, \citenamefont {Bloch}, \citenamefont {Eisert},\ and\ \citenamefont
  {Schneider}}]{braun2015emergence}%
  \BibitemOpen
  \bibfield  {author} {\bibinfo {author} {\bibfnamefont {S.}~\bibnamefont
  {Braun}}, \bibinfo {author} {\bibfnamefont {M.}~\bibnamefont {Friesdorf}},
  \bibinfo {author} {\bibfnamefont {S.~S.}\ \bibnamefont {Hodgman}}, \bibinfo
  {author} {\bibfnamefont {M.}~\bibnamefont {Schreiber}}, \bibinfo {author}
  {\bibfnamefont {J.~P.}\ \bibnamefont {Ronzheimer}}, \bibinfo {author}
  {\bibfnamefont {A.}~\bibnamefont {Riera}}, \bibinfo {author} {\bibfnamefont
  {M.}~\bibnamefont {Del~Rey}}, \bibinfo {author} {\bibfnamefont
  {I.}~\bibnamefont {Bloch}}, \bibinfo {author} {\bibfnamefont
  {J.}~\bibnamefont {Eisert}}, \ and\ \bibinfo {author} {\bibfnamefont
  {U.}~\bibnamefont {Schneider}},\ }\href@noop {} {\bibfield  {journal}
  {\bibinfo  {journal} {Proceedings of the National Academy of Sciences}\
  }\textbf {\bibinfo {volume} {112}},\ \bibinfo {pages} {3641} (\bibinfo {year}
  {2015})}\BibitemShut {NoStop}%
\bibitem [{\citenamefont {Anquez}\ \emph {et~al.}(2016)\citenamefont {Anquez},
  \citenamefont {Robbins}, \citenamefont {Bharath}, \citenamefont
  {Boguslawski}, \citenamefont {Hoang},\ and\ \citenamefont
  {Chapman}}]{anquez2016quantum}%
  \BibitemOpen
  \bibfield  {author} {\bibinfo {author} {\bibfnamefont {M.}~\bibnamefont
  {Anquez}}, \bibinfo {author} {\bibfnamefont {B.}~\bibnamefont {Robbins}},
  \bibinfo {author} {\bibfnamefont {H.}~\bibnamefont {Bharath}}, \bibinfo
  {author} {\bibfnamefont {M.}~\bibnamefont {Boguslawski}}, \bibinfo {author}
  {\bibfnamefont {T.}~\bibnamefont {Hoang}}, \ and\ \bibinfo {author}
  {\bibfnamefont {M.}~\bibnamefont {Chapman}},\ }\href@noop {} {\bibfield
  {journal} {\bibinfo  {journal} {Physical review letters}\ }\textbf {\bibinfo
  {volume} {116}},\ \bibinfo {pages} {155301} (\bibinfo {year}
  {2016})}\BibitemShut {NoStop}%
\bibitem [{\citenamefont {Gardas}\ \emph {et~al.}(2018)\citenamefont {Gardas},
  \citenamefont {Dziarmaga}, \citenamefont {Zurek},\ and\ \citenamefont
  {Zwolak}}]{gardas2018defects}%
  \BibitemOpen
  \bibfield  {author} {\bibinfo {author} {\bibfnamefont {B.}~\bibnamefont
  {Gardas}}, \bibinfo {author} {\bibfnamefont {J.}~\bibnamefont {Dziarmaga}},
  \bibinfo {author} {\bibfnamefont {W.~H.}\ \bibnamefont {Zurek}}, \ and\
  \bibinfo {author} {\bibfnamefont {M.}~\bibnamefont {Zwolak}},\ }\href@noop {}
  {\bibfield  {journal} {\bibinfo  {journal} {Scientific reports}\ }\textbf
  {\bibinfo {volume} {8}},\ \bibinfo {pages} {4539} (\bibinfo {year}
  {2018})}\BibitemShut {NoStop}%
\bibitem [{\citenamefont {Keesling}\ \emph {et~al.}(2018)\citenamefont
  {Keesling}, \citenamefont {Omran}, \citenamefont {Levine}, \citenamefont
  {Bernien}, \citenamefont {Pichler}, \citenamefont {Choi}, \citenamefont
  {Samajdar}, \citenamefont {Schwartz}, \citenamefont {Silvi}, \citenamefont
  {Sachdev} \emph {et~al.}}]{keesling2018probing}%
  \BibitemOpen
  \bibfield  {author} {\bibinfo {author} {\bibfnamefont {A.}~\bibnamefont
  {Keesling}}, \bibinfo {author} {\bibfnamefont {A.}~\bibnamefont {Omran}},
  \bibinfo {author} {\bibfnamefont {H.}~\bibnamefont {Levine}}, \bibinfo
  {author} {\bibfnamefont {H.}~\bibnamefont {Bernien}}, \bibinfo {author}
  {\bibfnamefont {H.}~\bibnamefont {Pichler}}, \bibinfo {author} {\bibfnamefont
  {S.}~\bibnamefont {Choi}}, \bibinfo {author} {\bibfnamefont {R.}~\bibnamefont
  {Samajdar}}, \bibinfo {author} {\bibfnamefont {S.}~\bibnamefont {Schwartz}},
  \bibinfo {author} {\bibfnamefont {P.}~\bibnamefont {Silvi}}, \bibinfo
  {author} {\bibfnamefont {S.}~\bibnamefont {Sachdev}},  \emph {et~al.},\
  }\href@noop {} {\bibfield  {journal} {\bibinfo  {journal} {arXiv preprint
  arXiv:1809.05540}\ } (\bibinfo {year} {2018})}\BibitemShut {NoStop}%
\bibitem [{\citenamefont {Xu}\ \emph {et~al.}(2014)\citenamefont {Xu},
  \citenamefont {Han}, \citenamefont {Sun}, \citenamefont {Xu}, \citenamefont
  {Tang}, \citenamefont {Li},\ and\ \citenamefont {Guo}}]{xu2014quantum}%
  \BibitemOpen
  \bibfield  {author} {\bibinfo {author} {\bibfnamefont {X.-Y.}\ \bibnamefont
  {Xu}}, \bibinfo {author} {\bibfnamefont {Y.-J.}\ \bibnamefont {Han}},
  \bibinfo {author} {\bibfnamefont {K.}~\bibnamefont {Sun}}, \bibinfo {author}
  {\bibfnamefont {J.-S.}\ \bibnamefont {Xu}}, \bibinfo {author} {\bibfnamefont
  {J.-S.}\ \bibnamefont {Tang}}, \bibinfo {author} {\bibfnamefont {C.-F.}\
  \bibnamefont {Li}}, \ and\ \bibinfo {author} {\bibfnamefont {G.-C.}\
  \bibnamefont {Guo}},\ }\href@noop {} {\bibfield  {journal} {\bibinfo
  {journal} {Physical review letters}\ }\textbf {\bibinfo {volume} {112}},\
  \bibinfo {pages} {035701} (\bibinfo {year} {2014})}\BibitemShut {NoStop}%
\bibitem [{\citenamefont {Cui}(2016)}]{cui2016j}%
  \BibitemOpen
  \bibfield  {author} {\bibinfo {author} {\bibfnamefont {J.}~\bibnamefont
  {Cui}},\ }\href@noop {} {\bibfield  {journal} {\bibinfo  {journal} {Sci
  Rep.}\ }\textbf {\bibinfo {volume} {6}},\ \bibinfo {pages} {33381} (\bibinfo
  {year} {2016})}\BibitemShut {NoStop}%
\bibitem [{\citenamefont {Gong}\ \emph {et~al.}(2016)\citenamefont {Gong},
  \citenamefont {Wen},\ and\ \citenamefont {Sun}}]{gong2016m}%
  \BibitemOpen
  \bibfield  {author} {\bibinfo {author} {\bibfnamefont {M.}~\bibnamefont
  {Gong}}, \bibinfo {author} {\bibfnamefont {X.}~\bibnamefont {Wen}}, \ and\
  \bibinfo {author} {\bibfnamefont {G.}~\bibnamefont {Sun}},\ }\href@noop {}
  {\bibfield  {journal} {\bibinfo  {journal} {Sci. Rep.}\ }\textbf {\bibinfo
  {volume} {6}},\ \bibinfo {pages} {22667} (\bibinfo {year}
  {2016})}\BibitemShut {NoStop}%
\bibitem [{\citenamefont {Cui}\ \emph {et~al.}(2020)\citenamefont {Cui},
  \citenamefont {G{\'o}mez-Ruiz}, \citenamefont {Huang}, \citenamefont {Li},
  \citenamefont {Guo},\ and\ \citenamefont {del
  Campo}}]{cui2020experimentally}%
  \BibitemOpen
  \bibfield  {author} {\bibinfo {author} {\bibfnamefont {J.-M.}\ \bibnamefont
  {Cui}}, \bibinfo {author} {\bibfnamefont {F.~J.}\ \bibnamefont
  {G{\'o}mez-Ruiz}}, \bibinfo {author} {\bibfnamefont {Y.-F.}\ \bibnamefont
  {Huang}}, \bibinfo {author} {\bibfnamefont {C.-F.}\ \bibnamefont {Li}},
  \bibinfo {author} {\bibfnamefont {G.-C.}\ \bibnamefont {Guo}}, \ and\
  \bibinfo {author} {\bibfnamefont {A.}~\bibnamefont {del Campo}},\ }\href@noop
  {} {\bibfield  {journal} {\bibinfo  {journal} {Communications Physics}\
  }\textbf {\bibinfo {volume} {3}},\ \bibinfo {pages} {1} (\bibinfo {year}
  {2020})}\BibitemShut {NoStop}%
\bibitem [{\citenamefont {Griffin}\ \emph {et~al.}(2012)\citenamefont
  {Griffin}, \citenamefont {Lilienblum}, \citenamefont {Delaney}, \citenamefont
  {Kumagai}, \citenamefont {Fiebig},\ and\ \citenamefont
  {Spaldin}}]{griffin2012scaling}%
  \BibitemOpen
  \bibfield  {author} {\bibinfo {author} {\bibfnamefont {S.~M.}\ \bibnamefont
  {Griffin}}, \bibinfo {author} {\bibfnamefont {M.}~\bibnamefont {Lilienblum}},
  \bibinfo {author} {\bibfnamefont {K.~T.}\ \bibnamefont {Delaney}}, \bibinfo
  {author} {\bibfnamefont {Y.}~\bibnamefont {Kumagai}}, \bibinfo {author}
  {\bibfnamefont {M.}~\bibnamefont {Fiebig}}, \ and\ \bibinfo {author}
  {\bibfnamefont {N.~A.}\ \bibnamefont {Spaldin}},\ }\href@noop {} {\bibfield
  {journal} {\bibinfo  {journal} {Physical Review X}\ }\textbf {\bibinfo
  {volume} {2}},\ \bibinfo {pages} {041022} (\bibinfo {year}
  {2012})}\BibitemShut {NoStop}%
\bibitem [{\citenamefont {Patane}\ \emph {et~al.}(2008)\citenamefont {Patane},
  \citenamefont {Silva}, \citenamefont {Amico}, \citenamefont {Fazio},\ and\
  \citenamefont {Santoro}}]{patane2008adiabatic}%
  \BibitemOpen
  \bibfield  {author} {\bibinfo {author} {\bibfnamefont {D.}~\bibnamefont
  {Patane}}, \bibinfo {author} {\bibfnamefont {A.}~\bibnamefont {Silva}},
  \bibinfo {author} {\bibfnamefont {L.}~\bibnamefont {Amico}}, \bibinfo
  {author} {\bibfnamefont {R.}~\bibnamefont {Fazio}}, \ and\ \bibinfo {author}
  {\bibfnamefont {G.~E.}\ \bibnamefont {Santoro}},\ }\href@noop {} {\bibfield
  {journal} {\bibinfo  {journal} {Physical review letters}\ }\textbf {\bibinfo
  {volume} {101}},\ \bibinfo {pages} {175701} (\bibinfo {year}
  {2008})}\BibitemShut {NoStop}%
\bibitem [{\citenamefont {Nalbach}\ \emph {et~al.}(2015)\citenamefont
  {Nalbach}, \citenamefont {Vishveshwara},\ and\ \citenamefont
  {Clerk}}]{nalbach2015quantum}%
  \BibitemOpen
  \bibfield  {author} {\bibinfo {author} {\bibfnamefont {P.}~\bibnamefont
  {Nalbach}}, \bibinfo {author} {\bibfnamefont {S.}~\bibnamefont
  {Vishveshwara}}, \ and\ \bibinfo {author} {\bibfnamefont {A.~A.}\
  \bibnamefont {Clerk}},\ }\href@noop {} {\bibfield  {journal} {\bibinfo
  {journal} {Physical Review B}\ }\textbf {\bibinfo {volume} {92}},\ \bibinfo
  {pages} {014306} (\bibinfo {year} {2015})}\BibitemShut {NoStop}%
\bibitem [{\citenamefont {Gao}\ \emph {et~al.}(2017)\citenamefont {Gao},
  \citenamefont {Zhang}, \citenamefont {Yu},\ and\ \citenamefont
  {Zhu}}]{gao2017anti}%
  \BibitemOpen
  \bibfield  {author} {\bibinfo {author} {\bibfnamefont {Z.-P.}\ \bibnamefont
  {Gao}}, \bibinfo {author} {\bibfnamefont {D.-W.}\ \bibnamefont {Zhang}},
  \bibinfo {author} {\bibfnamefont {Y.}~\bibnamefont {Yu}}, \ and\ \bibinfo
  {author} {\bibfnamefont {S.-L.}\ \bibnamefont {Zhu}},\ }\href@noop {}
  {\bibfield  {journal} {\bibinfo  {journal} {Physical Review B}\ }\textbf
  {\bibinfo {volume} {95}},\ \bibinfo {pages} {224303} (\bibinfo {year}
  {2017})}\BibitemShut {NoStop}%
\bibitem [{\citenamefont {Puebla}\ \emph {et~al.}(2020)\citenamefont {Puebla},
  \citenamefont {Smirne}, \citenamefont {Huelga},\ and\ \citenamefont
  {Plenio}}]{puebla2020universal}%
  \BibitemOpen
  \bibfield  {author} {\bibinfo {author} {\bibfnamefont {R.}~\bibnamefont
  {Puebla}}, \bibinfo {author} {\bibfnamefont {A.}~\bibnamefont {Smirne}},
  \bibinfo {author} {\bibfnamefont {S.~F.}\ \bibnamefont {Huelga}}, \ and\
  \bibinfo {author} {\bibfnamefont {M.~B.}\ \bibnamefont {Plenio}},\
  }\href@noop {} {\bibfield  {journal} {\bibinfo  {journal} {Physical Review
  Letters}\ }\textbf {\bibinfo {volume} {124}},\ \bibinfo {pages} {230602}
  (\bibinfo {year} {2020})}\BibitemShut {NoStop}%
\bibitem [{\citenamefont {Liou}\ and\ \citenamefont
  {Yang}(2018)}]{liou2018quench}%
  \BibitemOpen
  \bibfield  {author} {\bibinfo {author} {\bibfnamefont {S.-F.}\ \bibnamefont
  {Liou}}\ and\ \bibinfo {author} {\bibfnamefont {K.}~\bibnamefont {Yang}},\
  }\href@noop {} {\bibfield  {journal} {\bibinfo  {journal} {Physical Review
  B}\ }\textbf {\bibinfo {volume} {97}},\ \bibinfo {pages} {235144} (\bibinfo
  {year} {2018})}\BibitemShut {NoStop}%
\bibitem [{\citenamefont {Mukherjee}(2007)}]{mukherjee2007v}%
  \BibitemOpen
  \bibfield  {author} {\bibinfo {author} {\bibfnamefont {V.}~\bibnamefont
  {Mukherjee}},\ }\href@noop {} {\bibfield  {journal} {\bibinfo  {journal}
  {Phys. Rev. B}\ }\textbf {\bibinfo {volume} {76}},\ \bibinfo {pages} {174303}
  (\bibinfo {year} {2007})}\BibitemShut {NoStop}%
\bibitem [{\citenamefont {Divakaran}\ \emph {et~al.}(2009)\citenamefont
  {Divakaran}, \citenamefont {Mukherjee}, \citenamefont {Dutta},\ and\
  \citenamefont {Sen}}]{divakaran2009defect}%
  \BibitemOpen
  \bibfield  {author} {\bibinfo {author} {\bibfnamefont {U.}~\bibnamefont
  {Divakaran}}, \bibinfo {author} {\bibfnamefont {V.}~\bibnamefont
  {Mukherjee}}, \bibinfo {author} {\bibfnamefont {A.}~\bibnamefont {Dutta}}, \
  and\ \bibinfo {author} {\bibfnamefont {D.}~\bibnamefont {Sen}},\ }\href@noop
  {} {\bibfield  {journal} {\bibinfo  {journal} {Journal of Statistical
  Mechanics: Theory and Experiment}\ }\textbf {\bibinfo {volume} {2009}},\
  \bibinfo {pages} {P02007} (\bibinfo {year} {2009})}\BibitemShut {NoStop}%
\bibitem [{\citenamefont {Divakaran}(2008)}]{divakaran2008u}%
  \BibitemOpen
  \bibfield  {author} {\bibinfo {author} {\bibfnamefont {U.}~\bibnamefont
  {Divakaran}},\ }\href@noop {} {\bibfield  {journal} {\bibinfo  {journal}
  {Phys. Rev. B}\ }\textbf {\bibinfo {volume} {78}},\ \bibinfo {pages} {144301}
  (\bibinfo {year} {2008})}\BibitemShut {NoStop}%
\bibitem [{\citenamefont {Lieb}\ \emph {et~al.}(1961)\citenamefont {Lieb},
  \citenamefont {Schultz},\ and\ \citenamefont {Mattis}}]{lieb1961two}%
  \BibitemOpen
  \bibfield  {author} {\bibinfo {author} {\bibfnamefont {E.}~\bibnamefont
  {Lieb}}, \bibinfo {author} {\bibfnamefont {T.}~\bibnamefont {Schultz}}, \
  and\ \bibinfo {author} {\bibfnamefont {D.}~\bibnamefont {Mattis}},\
  }\href@noop {} {\bibfield  {journal} {\bibinfo  {journal} {Annals of
  Physics}\ }\textbf {\bibinfo {volume} {16}},\ \bibinfo {pages} {407}
  (\bibinfo {year} {1961})}\BibitemShut {NoStop}%
\bibitem [{\citenamefont {Bunder}(1999)}]{bunder1999je}%
  \BibitemOpen
  \bibfield  {author} {\bibinfo {author} {\bibfnamefont {J.}~\bibnamefont
  {Bunder}},\ }\href@noop {} {\bibfield  {journal} {\bibinfo  {journal} {Phys.
  Rev. B}\ }\textbf {\bibinfo {volume} {60}},\ \bibinfo {pages} {344} (\bibinfo
  {year} {1999})}\BibitemShut {NoStop}%
\bibitem [{\citenamefont {Caneva}\ \emph {et~al.}(2007)\citenamefont {Caneva},
  \citenamefont {Fazio},\ and\ \citenamefont {Santoro}}]{caneva2007adiabatic}%
  \BibitemOpen
  \bibfield  {author} {\bibinfo {author} {\bibfnamefont {T.}~\bibnamefont
  {Caneva}}, \bibinfo {author} {\bibfnamefont {R.}~\bibnamefont {Fazio}}, \
  and\ \bibinfo {author} {\bibfnamefont {G.~E.}\ \bibnamefont {Santoro}},\
  }\href@noop {} {\bibfield  {journal} {\bibinfo  {journal} {Physical Review
  B}\ }\textbf {\bibinfo {volume} {76}},\ \bibinfo {pages} {144427} (\bibinfo
  {year} {2007})}\BibitemShut {NoStop}%
\bibitem [{\citenamefont {Dziarmaga}(2010)}]{dziarmaga2010dynamics}%
  \BibitemOpen
  \bibfield  {author} {\bibinfo {author} {\bibfnamefont {J.}~\bibnamefont
  {Dziarmaga}},\ }\href@noop {} {\bibfield  {journal} {\bibinfo  {journal}
  {Advances in Physics}\ }\textbf {\bibinfo {volume} {59}},\ \bibinfo {pages}
  {1063} (\bibinfo {year} {2010})}\BibitemShut {NoStop}%
\bibitem [{\citenamefont {Dziarmaga}(2005)}]{dziarmaga2005dynamics}%
  \BibitemOpen
  \bibfield  {author} {\bibinfo {author} {\bibfnamefont {J.}~\bibnamefont
  {Dziarmaga}},\ }\href@noop {} {\bibfield  {journal} {\bibinfo  {journal}
  {Physical review letters}\ }\textbf {\bibinfo {volume} {95}},\ \bibinfo
  {pages} {245701} (\bibinfo {year} {2005})}\BibitemShut {NoStop}%
\bibitem [{\citenamefont {Zurek}\ \emph {et~al.}(2005)\citenamefont {Zurek},
  \citenamefont {Dorner},\ and\ \citenamefont {Zoller}}]{zurek2005dynamics}%
  \BibitemOpen
  \bibfield  {author} {\bibinfo {author} {\bibfnamefont {W.~H.}\ \bibnamefont
  {Zurek}}, \bibinfo {author} {\bibfnamefont {U.}~\bibnamefont {Dorner}}, \
  and\ \bibinfo {author} {\bibfnamefont {P.}~\bibnamefont {Zoller}},\
  }\href@noop {} {\bibfield  {journal} {\bibinfo  {journal} {Physical review
  letters}\ }\textbf {\bibinfo {volume} {95}},\ \bibinfo {pages} {105701}
  (\bibinfo {year} {2005})}\BibitemShut {NoStop}%
\bibitem [{\citenamefont {Mukherjee}\ \emph {et~al.}(2007)\citenamefont
  {Mukherjee}, \citenamefont {Divakaran}, \citenamefont {Dutta},\ and\
  \citenamefont {Sen}}]{mukherjee2007quenching}%
  \BibitemOpen
  \bibfield  {author} {\bibinfo {author} {\bibfnamefont {V.}~\bibnamefont
  {Mukherjee}}, \bibinfo {author} {\bibfnamefont {U.}~\bibnamefont
  {Divakaran}}, \bibinfo {author} {\bibfnamefont {A.}~\bibnamefont {Dutta}}, \
  and\ \bibinfo {author} {\bibfnamefont {D.}~\bibnamefont {Sen}},\ }\href@noop
  {} {\bibfield  {journal} {\bibinfo  {journal} {Physical Review B}\ }\textbf
  {\bibinfo {volume} {76}},\ \bibinfo {pages} {174303} (\bibinfo {year}
  {2007})}\BibitemShut {NoStop}%
\bibitem [{\citenamefont {Divakaran}\ \emph {et~al.}(2008)\citenamefont
  {Divakaran}, \citenamefont {Dutta},\ and\ \citenamefont
  {Sen}}]{divakaran2008quenching}%
  \BibitemOpen
  \bibfield  {author} {\bibinfo {author} {\bibfnamefont {U.}~\bibnamefont
  {Divakaran}}, \bibinfo {author} {\bibfnamefont {A.}~\bibnamefont {Dutta}}, \
  and\ \bibinfo {author} {\bibfnamefont {D.}~\bibnamefont {Sen}},\ }\href@noop
  {} {\bibfield  {journal} {\bibinfo  {journal} {Physical Review B}\ }\textbf
  {\bibinfo {volume} {78}},\ \bibinfo {pages} {144301} (\bibinfo {year}
  {2008})}\BibitemShut {NoStop}%
\end{thebibliography}%


\begin{thebibliography}{1}%
\makeatletter
\providecommand \@ifxundefined [1]{%
 \@ifx{#1\undefined}
}%
\providecommand \@ifnum [1]{%
 \ifnum #1\expandafter \@firstoftwo
 \else \expandafter \@secondoftwo
 \fi
}%
\providecommand \@ifx [1]{%
 \ifx #1\expandafter \@firstoftwo
 \else \expandafter \@secondoftwo
 \fi
}%
\providecommand \natexlab [1]{#1}%
\providecommand \enquote  [1]{``#1''}%
\providecommand \bibnamefont  [1]{#1}%
\providecommand \bibfnamefont [1]{#1}%
\providecommand \citenamefont [1]{#1}%
\providecommand \href@noop [0]{\@secondoftwo}%
\providecommand \href [0]{\begingroup \@sanitize@url \@href}%
\providecommand \@href[1]{\@@startlink{#1}\@@href}%
\providecommand \@@href[1]{\endgroup#1\@@endlink}%
\providecommand \@sanitize@url [0]{\catcode `\\12\catcode `\$12\catcode
  `\&12\catcode `\#12\catcode `\^12\catcode `\_12\catcode `\%12\relax}%
\providecommand \@@startlink[1]{}%
\providecommand \@@endlink[0]{}%
\providecommand \url  [0]{\begingroup\@sanitize@url \@url }%
\providecommand \@url [1]{\endgroup\@href {#1}{\urlprefix }}%
\providecommand \urlprefix  [0]{URL }%
\providecommand \Eprint [0]{\href }%
\providecommand \doibase [0]{http://dx.doi.org/}%
\providecommand \selectlanguage [0]{\@gobble}%
\providecommand \bibinfo  [0]{\@secondoftwo}%
\providecommand \bibfield  [0]{\@secondoftwo}%
\providecommand \translation [1]{[#1]}%
\providecommand \BibitemOpen [0]{}%
\providecommand \bibitemStop [0]{}%
\providecommand \bibitemNoStop [0]{.\EOS\space}%
\providecommand \EOS [0]{\spacefactor3000\relax}%
\providecommand \BibitemShut  [1]{\csname bibitem#1\endcsname}%
\let\auto@bib@innerbib\@empty
\bibitem [{\citenamefont
  {Khintchine}(1934)}]{khintchine1934korrelationstheorie}%
  \BibitemOpen
  \bibfield  {author} {\bibinfo {author} {\bibfnamefont {A.}~\bibnamefont
  {Khintchine}},\ }\href@noop {} {\bibfield  {journal} {\bibinfo  {journal}
  {Mathematische Annalen}\ }\textbf {\bibinfo {volume} {109}},\ \bibinfo
  {pages} {604} (\bibinfo {year} {1934})}\BibitemShut {NoStop}%
\end{thebibliography}%

\end{document}


\preprint{version 3.0 by amz}
\title{Supplementary material for ``Experimentally verifying anti-Kibble-Zurek
behavior in a quantum system under noisy control field''}
\author{Ming-Zhong Ai}
\author{Jin-Ming Cui}
\email{jmcui@ustc.edu.cn}

\author{Ran He}
\author{Zhong-Hua Qian}
\author{Xin-Xia Gao}
\author{Yun-Feng Huang}
\email{hyf@ustc.edu.cn}

\author{Chuan-Feng Li}
\email{cfli@ustc.edu.cn}

\author{Guang-Can Guo}
\affiliation{CAS Key Laboratory of Quantum Information, University of Science and
Technology of China, Hefei, 230026, People's Republic of China.}
\affiliation{CAS Center For Excellence in Quantum Information and Quantum Physics,
University of Science and Technology of China, Hefei, 230026, People's
Republic of China.}
\maketitle

\section{Gaussian noise}

\begin{figure}
\includegraphics{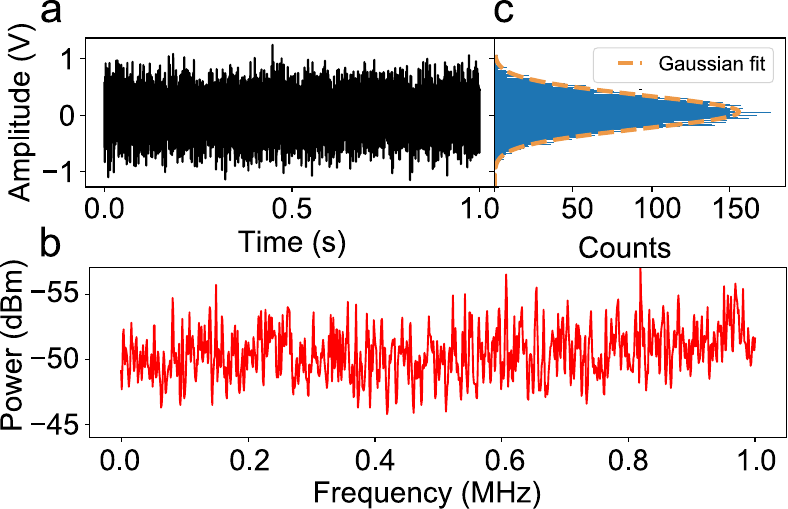}\caption{\label{fig:} (color online) The detailed information about white
Gaussian noise generated from microwave signal generator. (a) The
time domain information about these noise. (c) The corresponding histogram
of noise amplitudes in (a). We fit this histogram with a Gaussian
formula and the parameters of result is $\mu=-6.27{\rm e}-3,\sigma^{2}=0.1$.
(b) The frequency domain information of white Gaussian noise collected
with spectrum analyzer. The spectral density is a constant up to about
1 MHz, which indicates this noise is white noise.}
\end{figure}

The white Gaussian noise is a basic noise model in information theory
to mimic the effect of many random processes that occur in nature.
The white Gaussian noise used in our experiments is generated from
the microwave signal generator (E8257D and SG384) with a constant
spectral density and a Gaussian distribution. To verify the form of
this noise, we sample the white Gaussian noise from these devices
and the processed results are shown in Fig. \ref{fig:}.

\section{mapping intensity of noise to modulation parameter}

Suppose $\eta(t)$ is white Gaussian noise with zero mean and the
second moment $\left\langle \eta(t)\eta(\tau)\right\rangle =W^{2}\delta(\tau)$,
where the parameter $W^{2}$ represents the intensity of the noise.
According to Wiener-Khinchin-Kolmogorov theorem \citep{khintchine1934korrelationstheorie},
the power spectral density (PSD) of this noise is the Fourier transform
of its auto-correlation. That is
\begin{equation}
\left\langle \eta(t)\eta(\tau)\right\rangle =\int\eta(t)\eta(t-\tau)dt=\int\frac{d\omega}{2\pi}\left|g(\omega)\right|^{2}{\rm e}^{-i\omega t},
\end{equation}
in which $g(\omega)$ is the power spectral density. As for white
Gaussian noise, its PSD is a constant and $\delta(\tau)=\frac{1}{2\pi}\int{\rm e}^{i\omega\tau}d\omega$,
we can draw the conclusion that $g(\omega)=W$.

For the first quench protocol, the white Gaussian noise is add to
$\sigma_{z}$ part of the Hamiltonian and the noisy Hamiltonian $V^{(1)}=-2\eta(t)\sigma_{z}$.
Consider frequency modulation a cosine signal with white Gaussian
noise,

\begin{equation}
y_{1}(t)={\rm cos}((\omega_{res}+FG(t))t)
\end{equation}
where $\omega_{res}$ is the resonant frequency of internal state
of the ion, $F$ is the modulation deviation and the dimensionless
$G(t)$ is the noise signal. In the standard Hamiltonian of a two
level system of a ion, the detuning frequency $\Delta=FG(t)\propto\eta(t)$.
The PSD of this noise if the Fourier transform of its auto-correlation:

\begin{equation}
g(\omega)=\int_{0}^{1{\rm MHz}}dt\eta(t){\rm e}^{i\omega t}\label{eq:psd}
\end{equation}
. The bandwidth of this white Gaussian noise is up to 1 MHz, which
is far exceeds the Rabi frequency (about 10 kHz in our experiments).
Then we could conclude that $F\propto W$ and changing the modulation
deviation equals to changing the intensity of noise.

For the third quench protocol, the white Gaussian noise is add to
$\sigma_{x}$ part of the Hamiltonian and the noisy Hamiltonian $V^{3}=-2\eta(t)J{\rm sin}k\sigma_{x}$.
Amplitude modulation a cosine signal with white Gaussian noise can
be written as

\begin{equation}
y_{2}(t)=[1+AG(t)]{\rm cos}(\omega_{res}t),
\end{equation}
in which $A$ is the modulation depth. In the Hamiltonian of a ion,
the Rabi frequency $\Omega\propto AG(t)$ and that is to say $AG(t)\propto\eta(t)$.
According to Eq. \ref{eq:psd}, the final strength of noise is proportional
to modulation depth $A$.

As for the second quench protocol, the white Gaussian noise is add
to both $\sigma_{x}$ and $\sigma_{z}$ part of Hamiltonian. So frequency
modulation and amplitude modulation are necessary simultaneously.

\bibliographystyle{apsrev4-1}
\bibliography{SM-antiKZM}